# Picosecond Switching of Optomagnetic Tunnel Junctions

**Authors:**

Luding Wang,[1,2] Houyi Cheng,[1] Pingzhi Li,[2] Yang Liu,[1] Youri L. W. van Hees,[2] Reinoud Lavrijsen,[2] Xiaoyang Lin,[1] Kaihua Cao,[1] Bert Koopmans,[2] and Weisheng Zhao[1]

**Affiliations:**

[1]Fert Beijing Institute, School of Microelectronics, Beijing Advanced Innovation Center for Big Data and Brain Computing, Beihang University, 100191 Beijing, China

[2]Department of Applied Physics, Institute for Photonic Integration, Eindhoven University of Technology, P.O. Box 513, 5600 MB Eindhoven, The Netherland

* Author to whom correspondence should be addressed: weisheng.zhao@buaa.edu.cn

**Introductory paragraph**

Perpendicular magnetic tunnel junctions are one of the building blocks for spintronic memories, which allow fast nonvolatile data access, offering substantial potentials to revolutionize the mainstream computing architecture[1-3]. However, conventional switching mechanisms of such devices are fundamentally hindered by spin polarized currents[4], either spin transfer torque[5,6] or spin orbit torque[7,8] with spin precession time limitation and excessive power dissipation. These physical constraints significantly stimulate the advancement of modern spintronics[1,4]. Here, we report an optomagnetic tunnel junction using a spintronic-photonic combination. This composite device incorporates an all-optically switchable Co/Gd bilayer coupled to a CoFeB/MgO-based perpendicular magnetic tunnel junction by the Ruderman-Kittel-Kasuya-Yosida interaction. A picosecond all-optical operation of the optomagnetic tunnel junction is explicitly confirmed by time-resolved measurements. Moreover, the device shows a considerable tunnel magnetoresistance and thermal stability. This proof-of-concept device represents an essential step towards ultrafast spintronic memories with THz data access, as well as ultralow power consumption.

**Main text**

For decades, CoFeB/MgO-based perpendicular magnetic tunnel junctions (p-MTJs)[4,9] have become prominent components of magnetic random-access memory (MRAM), highly promising towards next-generation spintronic computing with non-von Neumann architecture[1,2]. As to the operation mechanisms of the p-MTJ bit cell, the prototype MRAMs typically relied on an external magnetic field, whereas state-of-the-art MRAM technologies are dedicated to spin polarized current-induced schemes[4-8]. Unfortunately, until today, the operation speed of these p-MTJs are still limited to sub-nanosecond whilst requiring extremely high current densities, reducing of which remained a long-existing vital challenge for modern spintronic R&D[1,10-12].

To address this issue, integrating all-optical switching (AOS)[13-15] with p-MTJs has received considerable interest[16] ever since AOS first discovered in ferrimagnetic GdFeCo alloys[13]. AOS offers emerging prospects towards picosecond operation of the p-MTJs, which is 1 to 2 orders of magnitude faster than conventional schemes, while offering an enhanced energy efficiency as well. Although having undergone inspiring research progress[17,18], until now, a fully functional all-optical operation of p-MTJ, with high tunnel magnetoresistance (TMR) and thermal stability are still missing. Furthermore, the operation speed of the OMTJ, along with its time-resolved switching dynamics, remains elusive.

Recently, deterministic AOS has been observed in synthetic ferrimagnetic Co/Gd bilayers[19,20]. Apart from the efficient and robust single-pulse AOS[21,22], they show extended flexibility on interface engineering, such as high domain wall (DW) velocity, and inherent built-in interfacial Dzyaloshinskii-Moriya interaction (iDMI)[23], which are indispensable components for future ultra-high density spintronic memories[24,25]. Moreover, recent study[26] shows that the AOS efficiency is considerably enhanced after 300°C thermal annealing, thus offering an ideal candidate for CoFeB/MgO-based p-MTJ integration due to a fully compatible fabrication process like post-annealing.

In this work, we design an optomagnetic tunnel junction (OMTJ) memory device. By a careful stack design, a composite structure, with an all-optically switchable Co/Gd bilayer coupled to a CoFeB layer by Ruderman-Kittel-Kasuya-Yosida (RKKY) interaction, is synthesized. The proposed OMTJ exhibits an efficient toggle AOS, as well as a reliable electrical TMR readout in real time. Lastly, the picosecond switching speed of the OMTJ device is explicitly confirmed by time-resolved magnetization dynamics.

As illustrated in Fig. 1(a), the designed OMTJ stack is composed of, from the substrate side upwards, Ta (3)/Ru (20)/Ta (0.7)/[Co/Pt]$_m$ (9.8)/Ru (0.8)/[Co/Pt]$_n$ (3.6)/Ta (0.3)/CoFeB (1.2)/MgO/CoFeB (1)/Ta (0.3)/Co (1)/Gd (3)/Pt (2) (numbers in parentheses denote the layer nominal thicknesses in nanometer, and the subscripts are the repeated numbers), which is deposited by DC and RF magnetron sputtering (see

Methods). In this configuration, Ta/Ru/Ta is the bottom electrode, [Co/Pt]$_{m/n}$ multilayers with the bottom CoFeB are the synthetic antiferromagnetic (SAF) reference layer (RL), whereas the MgO layer is the tunnel barrier. As to the synthetic free layer (FL), the top CoFeB layer and the Co/Gd bilayer are RKKY coupled through an atom-thick Ta spacer. After deposition, the stacks are subjected to thermal annealing at 300°C in vacuum for an hour.

Figure 1(b) shows the out-of-plane magnetization versus magnetic field (*M-H*) hysteresis loop of the OMTJ stack using a VSM-SQUID at room temperature (see Methods). The film exhibits the presence of strong perpendicular magnetic anisotropy (PMA) in both the FL and the RL, as indicated by the 100% remanence and squareness of the hysteresis loops. According to the steep minor loop (inset of Fig. 1(b)), two parts of the FLs switch simultaneously, due to the effective RKKY coupling through the thin Ta spacer, which is paramount for the OMTJ performance. Note that the coercive field of the FL reaches 4 mT, and a relatively small shift of the hysteresis loop indicates a significant reduction of stray field due to the SAF layers. The thermal stability of the OMTJ ranging from 250 K to 350 K, is measured by temperature dependence of *M-H* loops (Supplementary Information Note I).

We investigate the single-pulse AOS of the OMTJ full stack. In the measurements (see Methods), the sample is exposed to subsequent linearly polarized laser pulses, whereafter its magnetization response is measured using magneto-optical Kerr microscopy in a static state. As shown in the Fig. 1(c), five separate spots are excited by a different number of laser pulses (Row I). We observe that a homogeneous domain with an opposite magnetization direction is written for every odd number of laser pulses, whereas it toggles back for every even number of pulses. Then, the OMTJ stack is exposed to a laser-pulse train with partly overlapping area (Row II - III, see Methods). As to the regions exposed by a single or three-overlapping pulses, a reversed domain is observed. In contrast, for the regions where only two pulses overlap, no net reversal is observed. Here, we also note that the DW at the overlapping regions stay intact, whereas the domains toggle upon every laser excitation. The observed results are consistent with the thermal AOS mechanism discussed in previous studies[14,19,20], demonstrating a robust toggle AOS in the OMTJ stack.

As to the next-generation memory paradigm, it is always essential to reduce the energy needed to switch a memory bit cell. Therefore, the AOS efficiency of the OMTJ stack is investigated, by measuring the pulse energy dependence of the domain size (see Methods). As shown in Fig. 1(d), a threshold laser energy ($P_0$) of 330 nJ is observed. Above $P_0$, the laser written domain size increases with higher laser energies. The AOS threshold fluence ($F_0$) of 3.1 mJ/cm$^2$ is then determined by assuming a Gaussian shape of the laser pulse[19]. The observed $F_0$ is relatively lower than that for GdFeCo alloys[13] or (Co/Tb)$_n$ based electrodes[18]. Further reduction of $F_0$ can be expected by decreasing the thickness of the ferromagnetic layer, as shown in previous studies[19,21]. Moreover, due to the thermal nature of the AOS, the energy for writing a sub-20 nm OMTJ bit could scale down to a few tens of fJs using a plasmonic antenna[21,27] as already used in heat assisted magnetic recording (HAMR) techniques, which is potentially competitive with state-of-the-art spintronic memories[16].

Circular pillars are patterned using multistep UV lithography and an Ar ion milling process (see Methods). Figure 2(a) shows an optical microscope image of the fabricated OMTJ device with a pillar diameter of 3 μm (see the inset of Fig. 2(a) for the zoom-in of the pillar) and four electrode pads. Note that the 100-nm-thick indium tin oxide (ITO) is employed as top electrodes, which is crucial for our hybrid optospintronic device. Compared with the conventional Ti/Au electrode, the transparent ITO enables an efficient laser-pulse access, as well as a reliable electrical detection with a high signal-noise ratio (SNR). The resistance versus magnetic field (*R-H*) loop is measured by sweeping an out-of-plane magnetic field, as shown in Fig. 2(b). A clear bi-stable tunnel magnetoresistance (TMR) is observed, with $R_{AP} = 227\ \Omega$ (AP: antiparallel) and $R_P = 169\ \Omega$ (P: parallel), respectively. The RKKY coupled FL is found to switch as a single unit, as indicated by steep resistance transition events at +190 mT and −115 mT, respectively, where the slight bias field indicates an effective reduction of the stray field due to the RL. A typical TMR ratio $(R_{Ap} - R_P)/R_P = 34\%$ is obtained for our proof-of-concept OMTJ device after post-annealing, which could be further enhanced by optimizing the stack design, but goes beyond the aim of the present work. In addition, the improved SNR in the *R-H* measurement indicates an improved interface quality between the ITO and the OMTJ pillar.

Next, we investigate the all-optical operation of the OMTJ device. As sketched in Fig. 2(c), the programming of the OMTJ is demonstrated using subsequent laser pulses, whereas the read operation is realized by an electrical TMR measurement in real time (see Methods). The complete measurement is performed without any external magnetic field. As shown in Fig. 2(d), the TMR of the OMTJ device toggles deterministically between the AP and the P state at the same frequency as the incoming laser pulses. In other words, programming a "1"/ "0" is realized for every odd/even number of pulses. The reconfigurability of the operation is further verified by a 100% success rate up to millions of repeated tests. Moreover, the magnetoresistance values measured by AOS are equal to the ones in the *R-H* loop (Fig. 2(b)), unambiguously demonstrating a complete reversal of the composite FL in the bottom-pinned OMTJ. Lastly, we stress that the design rules of such OMTJs are not limited to Gd/Co bilayers. Other ferrimagnetic AOS system, such as GdFeCo or (Co/Tb)$_n$, also hold great potential for further optimized performance, already with some highly inspiring progress[17,18].

Lastly, we investigate the operation speed of the OMTJ. Previous studies[13,28] have reported tens of picosecond AOS in some ferrimagnetic alloys. However, no time-resolved AOS dynamics have been performed in an MTJ with additional magnetic layers yet. Moreover, in the RKKY coupled FL, whether the switching speed is impeded has also remained unclear. Thus, time-resolved magneto-optical Kerr (TR-MOKE) measurements are performed (see Methods) to clarify such concerns.

As illustrated in Fig. 3(a), the laser-induced magnetization dynamics of the OMTJ stack shows typical characteristics that are consistent with previous studies on time-resolved AOS studies[13,28,29]. Briefly, in the first several picosecond time scale, a rapid demagnetization is caused by the ultrafast laser heating. Afterwards, magnetization reversal takes place, due to the distinct demagnetization times[13] between the antiferromagnetically coupled Co and Gd sublattices, which is driven by the angular momentum transfer mediated by exchange scattering[20]. Lastly, a long-lasting tail up to hundreds of picoseconds indicates the magnetization settling to a new thermal equilibrium in the reversed orientation. More importantly, we observe the zero-crossing point (see inset of Fig. 3(a) for zoom-in) of several picoseconds, and the almost fully reversal time within tens of picoseconds. These results confirm that the CoFeB layer follows the Co/Gd quasi instantaneously on a picosecond time scale.

We investigate the operation speed of the patterned device, as shown in Fig. 3(b), with pillar diameters of 10 μm, 5 μm, 3 μm, respectively. Note that the relatively low SNR, compared to Fig. 3(a), is attributed to the OMTJ pillar size being smaller than the probe spot, reducing the total magnetic signal that we collect. We observe that the picosecond AOS speed is preserved upon scaling down to 3 μm, as proved by a zero-cross point also within 5 picoseconds. The operation speed of the OMTJ is here defined by the time ($t_{switch}$) needed to reach 75% opposite magnetization state, which is chosen because of an ample read margin for binary electronics chips[28]. As plotted in the inset of Fig. 4(b), the $t_{switch}$ of the OMTJ devices are in the range of tens of picoseconds. In addition, our results may even indicate a scaling dependence of $t_{switch}$, enhancing upon reduced dimensions. Although care has been taken because of the limited SNR, the trend in our results are in reasonable agreement with recent studies using GdCo nanodots[28]. Especially, compared to $t_{switch} \approx 40$ ps for the full-sheet stack, it reduces to 10 -- 25 ps for micro-sized OMTJ device. We conjuncture that the nonuniform heat diffusion to the ambient, that leads the magnetization settle to a new thermal equilibrium faster may play a crucial role in this observed size dependence. However, other laser-induced magnetization phenomena, such as spin-lattice coupling and energy transfer rates as speculated in previous studies[28], may not be fully ruled out. More insights regarding the downscaling of OMTJs will be obtained in future work. Above all, by time-resolved studies, we demonstrate a picosecond operation speed of the OMTJ device, highly promising towards future applications like THz opto-MRAM chips.

In conclusion, we report an integrated spintronic-photonic OMTJ memory device. The composite device incorporates an all-optically switchable Co/Gd bilayer that is RKKY coupled to a CoFeB/MgO-based p-MTJ. The ultrafast all-optical memory performance, with reliable electrical read operation, is experimentally demonstrated. The proposed device provides an essential milestone towards large-scale opto-MRAM arrays, within the scope of THz memory/computing architecture. Additionally, by direct memory of femtosecond optical information, it represents a new category of nonvolatile opto-memory, which extends the inherent advantage of photonics like data transfer and processing. Thus, the proof-of-concept OMTJ exhibits great potential to stimulate the innovation of future & emerging technologies (FET).

**Methods**

**Sample deposition.** The OMTJ stacks used in this work were deposited using DC and RF magnetron sputtering (AJA International Physical Vapor Deposition) at room temperature, which were deposited on a thermally oxidized Si (001) substrate at a base pressure in the deposition chamber of $10^{-9}$ mbar without an external magnetic field. The CoFeB target composition was $Co_{20}Fe_{60}B_{20}$ (in atomic percent), with a deposition rate of 3 min/nm at Ar pressure of $8*10^{-4}$ mbar. The deposition rate for MgO was 5.5 min/nm at Ar pressure of $8*10^{-4}$ mbar. After deposition, the stacks were annealed in vacuum (with a base pressure of $10^{-9}$ mbar) at 300°C for one hour without an external magnetic field.

**Device fabrication.** Micro-sized OMTJ pillars were patterned by using a standard UV optical lithography in combination with argon ion milling process, at the centre of Ta/Ru/Ta bottom electrode. The diameter of the OMTJ pillars were 10, 5, 4, 3 μm, respectively. The samples were then covered with $SiO_2$ insulation by electron beam evaporation (EBV) and a lift-off procedure. Subsequently, the 100-nm-thick transparent indium tin oxide (ITO) were deposited as the top electrodes, also by EBV and a standard lift-off procedure. The quality of ITO deposition process is essential for the OMTJ device, leading to an efficient laser-pulse access, and a reliable electrical detection as well.

**Magnetic properties characterization.** The magnetic characteristics of the full-sheet OMTJ stack were investigated at room temperature (Fig. 1(b)) using a vibrating sample magnetometer - superconducting quantum interference device (VSM-SQUID), under an out-of-plane magnetic field ranging from ±400 mT.

The tunnel magnetoresistance of the OMTJ were characterized at room temperature by a conventional four-point TMR measurement under an out-of-plane magnetic field in the range of ±200 mT. To measure the TMR signal, a small current (100 μA) was sent through the OMTJ device, whereas the resulting magnetoresistance was measured using a lock-in amplifier.

**Deterministic all-optical switching measurements.** The response of magnetization in the OMTJ stack upon subsequent femtosecond laser pulses (Spectra Physics Spirit-NOPA) were investigated. The laser pulse was linearly polarized, with a pulse duration of ≈ 100 fs at sample position, a central wavelength of 700 nm, a spot radius (1/e Gaussian pulse) typically of 25 μm, and a base repetition rate of 500 kHz. By using a pulse picker and a mechanical shutter, individual laser pulses could be picked out. In the single-pulse AOS measurements (Fig. 1(c), Row I), which were performed at room temperature, the magnetization was first saturated by an external field. Afterwards, the field was turned off and the stack was exposed to subsequent laser pulses. The numbers labelled in the figure correspond to the laser-pulse numbers of each spot. Then the OMTJ stack was exposed to laser-pulse trains with partly overlapping areas. The laser-pulse train was set at a velocity of 0.2 mm/s, with a repetition rate of 2.5 Hz (Row II) and 5 Hz (Row III), respectively.

The responses of the magnetization after laser-pulse excitation were measured in steady state using magneto-optical Kerr microscopy, where light and dark regions were corresponding to up and down magnetization direction. As to the Kerr microscopy images, a differential technique was used to enhance the magnetic contrast. Specifically, a "background" image was captured in the magnetization saturation state. This "background" was then subtracted from the subsequent Kerr images after laser-pulse excitation. The scale bars in the Kerr images were 200 μm.

To determine the AOS threshold fluence of the OMTJ stack, the OMTJ stack was exposed to single laser pulses with different laser energies ($E_P$) as indicated in the inset of Fig. 1(d) (in unit of nJ), whereafter the laser-pulse energy dependence of the AOS domain size was measured by Kerr microscopy (see inset of Fig. 1(d) for the Kerr microscopy image). By assuming a Gaussian energy profile of the laser pulse, the AOS threshold fluence could then be determined.

**All-optical operation of OMTJ device.** To investigate the all-optical programming operation of the device, the fabricated OMTJs were excited by subsequent linearly polarized fs laser pulses, with the same laser configuration described above. The laser pulse train was set at a relatively low repetition rate of 0.5 Hz to identify each single pulse. Meanwhile, as to the read operation, we measured its real-time electrical readout upon laser-pulse excitation, using a four-point TMR measurement. The complete measurements were performed without external magnetic field.

**Time-resolved Kerr measurements.** TR-MOKE measurements were performed using a typical pump-probe configuration at a repetition rate of 100 kHz. In the measurements, the sample was first exposed by a pump pulse with a duration of 100 fs, a spot size of typically 35 μm and a laser-pulse energy of typically 600 nJ to write an AOS domain. Meanwhile, a probe pulse, which arrived at the sample with a different time delay and much lower laser energy, measures the time evolution of the magnetization via the Magneto-optic Kerr effect. The probe spot size was typically 12 μm for successful detection for OMTJ pillars, integrating the magnetic signal over the full OMTJ pillar element. Due to the toggle switching behavior of the OMTJ, an external magnetic field with an opposite direction was applied, which is slightly higher than the sample's coercive filed. This method is consistent with other time-resolved AOS studies[28,29]. A series of TR-MOKE measurements were averaged by using both positive and negative magnetic field direction. Data normalization was done based on the averaged magnetization Kerr signals in both positive and negative saturation states, which was extracted from the averaged TR-MOKE results for each pillar size.

**Acknowledgments.**

We gratefully acknowledge the National Key R&D Program of China 2018YFB0407602, the National Natural Science Foundation of China (Grant No. 61627813), the International Collaboration Project B16001 and China Scholarship Council (CSC) for their financial support of this work. This work is also part of the Gravitation program "Research Centre for Integrated Nanophotonics," which was financed by the Netherlands Organisation for Scientific Research (NWO).

**Author contributions**.

W.Z. designed, initiated and supervised the project. L.W., H.C. and Y.L. deposited the stacks, fabricated the devices. L.W. initiated the measurements under the supervisor of B.K. P.L., Y.H. contributes to static and time-resolved all-optical switching setup & measurements. L.W., X.L., K.C. and R.L. analysed and discussed the results. L.W. and W.Z. drafted the manuscript and all authors reviewed the manuscript.

**Data and materials availability**.

All data supporting the findings of this study are available from the corresponding author upon reasonable request.

**Competing Interests**.

The authors declare no competing interests.

**Figures**

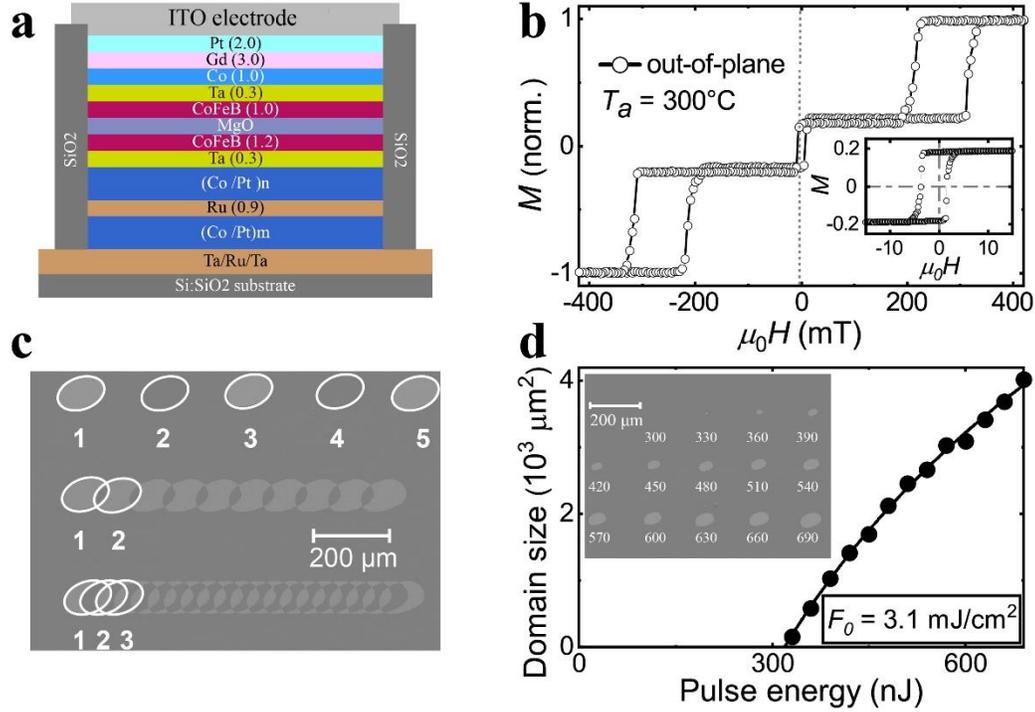

**Fig.1 OMTJ structure, magnetic characteristics, and efficient AOS of the stack. a.** Schematic structure of the proposed OMTJ memory device. Ta/Ru/Ta and ITO are the bottom and top electrode, respectively. [Co/Pt]$_{m/n}$ multilayers based SAF is used as the reference layer (RL), MgO as the tunnel barrier. The composite free layer (FL) consists of the top CoFeB layer and the Co/Gd bilayer, which are RKKY coupled through an atom-thick Ta spacer. **b.** Out-of-plane hysteresis loop of the OMTJ stack after post-annealing at 300°C measured by a VSM-SQUID. The square hysteresis loops with 100% remanence indicates a well-defined PMA in both FL and RL. The minor loop of the OMTJ stack, as shown in the inset, indicates that two parts of the FLs switch simultaneously due to the strong RKKY coupling. **c.** Deterministic single-pulse AOS measured by Kerr microscopy. Row (I): Single-pulse AOS measurement by subsequent fs laser pulses. The numbers correspond to the number of pulses that the region is exposed to. Row (II -- III): AOS measurements by laser pulse train at a repetition rate of 2.5 Hz and 5 Hz, respectively, leading to partly overlapping pulses. **d.** Laser-pulse energy dependence of the AOS domain size, showing the threshold laser energy ($P_0$) of 330 nJ, and an increasing domain size with higher laser energies. The threshold fluence ($F_0$) of the OMTJ stack is calculated to be 3.1 mJ/cm$^2$. Inset: Kerr microscopy of the measurement performed on the OMTJ stack. The numbers correspond to the laser energy of each spot in unit of nJ.

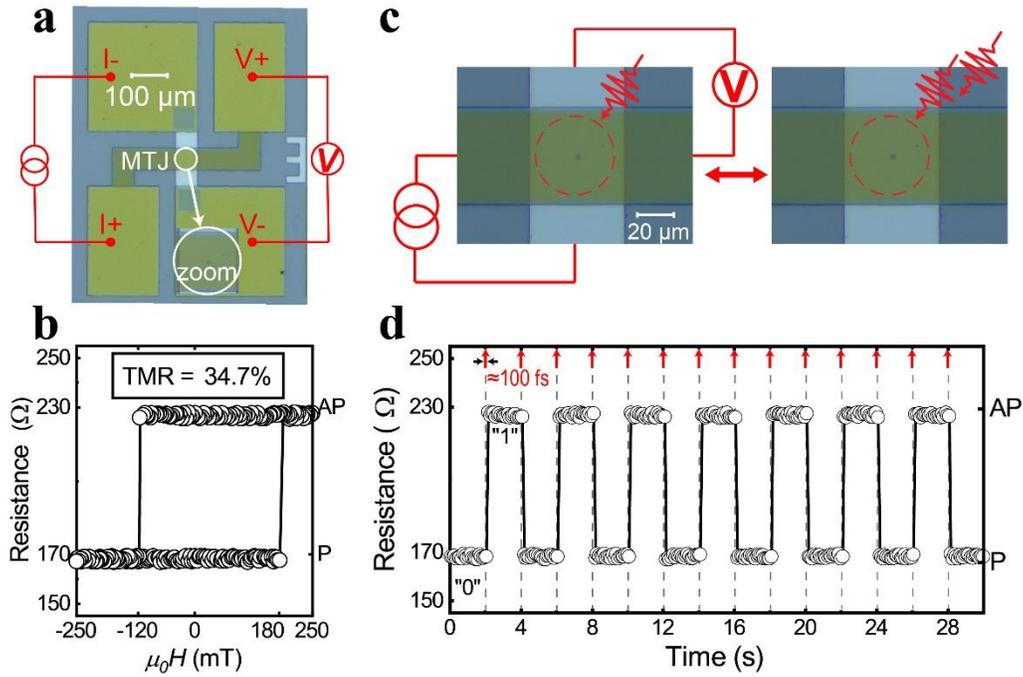

**Fig. 2 OMTJ device characterization, and all-optical "write" operation. a.** Microscope image of the fabricated OMTJ device with a 100-nm-thick transparent ITO top electrode, as well as four electrode pads to perform 4-point TMR detection. The inset shows the zoom-in of the OMTJ pillar. **b.** *R-H* magnetoresistance loop measured by sweeping an out-of-plane magnetic field, showing a typical TMR ratio of 34%. **c.** Schematic overview of the AOS "write" operation of the OMTJ device. A small current is applied through the OMTJ, while the resulting TMR voltage is measured in real time. The OMTJ pillar is excited by a train of linearly polarized laser pulses. **d.** Typical TMR measurement as a function of time upon laser-pulse excitation. The resistance toggles between P and AP state upon every laser pulse excitation. No external field was applied during the measurements.

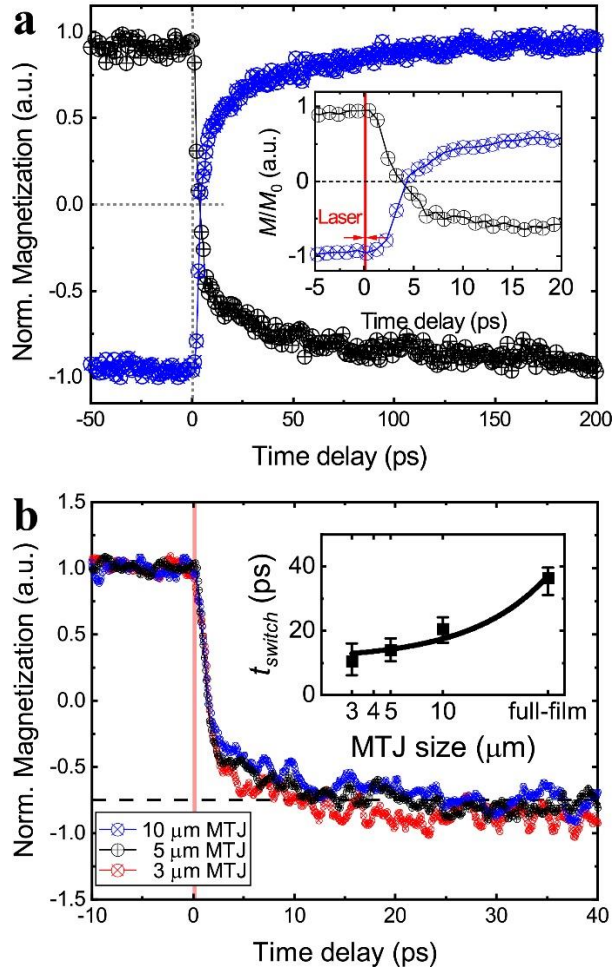

**Fig. 3 Time-resolved measurements of the OMTJ operation speed. a.** Time-resolved magneto-optical Kerr (TR-MOKE) measurements performed on the OMTJ full stack. The inset shows the zoom-in of the first 20 ps time scale. The switching speed is confirmed by the zero-crossing time below 5 picoseconds. **b.** TR-MOKE measurements performed on the OMTJ with pillar diameters of 10 μm, 5 μm, 3 μm, respectively. The results show that the ultrafast AOS speed is unimpeded in patterned devices, as proven by a zero-cross point also in the picosecond time scale. Inset: the time needed for 75% magnetization reversal ($t_{switch}$) as a function of OMTJ pillar size. The results may indicate a scaling dependence of $t_{switch}$, i.e., compared to $t_{switch} \approx 40$ ps for the full-sheet stack, it reduces to 10 to 25 picoseconds for the OMTJ device. Although care has been taken because of the limited SNR, the tendency is in reasonable agreement with recent studies using GdCo nanodots[28].